\documentclass[10pt, a4paper]{article}

\usepackage{lrec-coling2024} 
\usepackage{times}
\usepackage{latexsym}
\usepackage{graphicx}

\usepackage[T1]{fontenc}

\usepackage[utf8]{inputenc}

\usepackage{microtype}
\usepackage{inconsolata}
\usepackage{algorithm}
\usepackage{algorithmic}
\usepackage{multirow}
\usepackage{makecell}
\usepackage{array}
\usepackage{booktabs}
\usepackage{amsfonts}
\usepackage{amsmath}
\usepackage{tabularx}
\usepackage{enumitem}
\newcommand\italicRoman[1]{
    \textit{\expandafter{\romannumeral#1})}
}
\title{AntCritic: Argument Mining for Free-Form and Visually-Rich Financial Comments}

\name{Huadai Liu$^{1*}$\thanks{*Equal contributions}, Wenqiang Xu$^{2*}$, Xuan Lin$^{2*}$, Jingjing Huo$^{2}$, Hong Chen$^{2}$, Zhou Zhao$^{1}$} 

\address{Zhejiang University$^{1}$, Ant Group$^{2}$ \\
         liuhuadai@zju.edu.cn, \{yugong.xwq, daxuan.lx,huojingjing.hjj\}@antgroup.com \\
         wuyi.ch@antgroup.com, zhaozhou@zju.edu.cn\\}

\abstract{
Argument mining aims to detect all possible argumentative components and identify their relationships automatically. As a thriving task in natural language processing, there has been a large amount of corpus for academic study and application development in this field. However, the research in this area is still constrained by the inherent limitations of existing datasets. Specifically, all the publicly available datasets are relatively small in scale, and few of them provide information from other modalities to facilitate the learning process. Moreover, the statements and expressions in these corpora are usually in a \textit{compact} form, which restricts the generalization ability of models. To this end, we collect a novel dataset \textit{AntCritic} to serve as a helpful complement to this area, which consists of about 10k free-form and visually-rich financial comments and supports both argument component detection and argument relation prediction tasks. Besides, to cope with the challenges brought by scenario expansion, we thoroughly explore the fine-grained relation prediction and structure reconstruction scheme and discuss the encoding mechanism for visual styles and layouts. On this basis, we design two simple but effective model architectures and conduct various experiments on this dataset to provide benchmark performances as a reference and verify the practicability of our proposed architecture. 
We release our data and code in this \href{https://drive.google.com/drive/folders/1N26tuhMH2hJWoUrNQAZYoUkERB7OHsCS?usp=sharing}{\textit{link}}, and this dataset follows CC BY-NC-ND 4.0 license.
 \\ \newline  }

\begin{document}

\maketitleabstract

\section{Introduction}
With the rapid development of Internet technology, millions of opinions and thoughts are thus transmitted and stored on various web pages and applications, containing a great deal of research and application value. Among them, some complicated scenarios bring extra challenges, such as processing academic exchanges or movie reviews. To this end, the task of argument mining appears and arouses growing attention.


Given a persuasive article containing argumentative expressions, the task of argument mining aims to automatically identify all the potential argument components and correctly recover the corresponding argument structures. Unlike the other tasks, such as sentiment analysis or key-phrase extraction, this target requires the intelligent model to concurrently understand both semantic information and the logical structure and causal relationships within data. Despite the difficulties and challenges in the modeling process, the importance of this task is apparent. From the academic view, it can provide accurate information for the derivative subtasks (such as argument summarization and structured argument generation) and is conducive to understanding documents with complex semantic structures. While for the application aspect, the learned models can assist businesses or enterprises to automatically extract opinions and feedback or quickly trace the core opinions of prevailing topics without too much manual effort.

As a comprehensive and inclusive field, argument mining can be regarded as a combination of multiple interrelated sub-goals in essence. Therefore, to put more emphasis on academic analysis rather than constructing a complicated pipeline, researchers tend to discuss and address the issues only in a specific section, such as argument component detection \cite{Gao2017ReinforcementLB, Daxenberger2017WhatIT, Ruggeri2021TreeConstrainedGN}, argument relation classification \cite{Nguyen2016ContextawareAR, Peldszus2015TowardsDC, Cocarascu2018CombiningDL}, and stance detection \cite{Wei2018MultiTargetSD, Ebrahimi2016WeaklyST, Johnson2016IdentifyingSB}. 
And to facilitate these subtasks, massive language resources \cite{Reed2008LanguageRF, Habernal2017ArgumentationMI, Peldszus2015AnAC, Stab2017ParsingAS, Fergadis2021ArgumentationMI} are devoted to providing more effective and diversified data and supporting the learning process of models. Although the sufficient research and corpus mentioned above have established the foundation of this task, there is still a lack of further analysis for more variable and challenging scenarios. For example, the previous work tends to assume that all the arguments are expressed in a \textit{compact} form, which means the statements of a single unit will be arranged consecutively. Thus non-adjacent clauses or sentences will be naturally regarded as different ones. But in practice, there might be some situations where semantic-consistent claims and premises get entwined with each other. Besides, although a considerable number of visually-rich documents are collected from social media such as Twitter \cite{Bosc2016DARTAD}, Wikipedia \cite{Biran2011IdentifyingJI}, and web blog posts \cite{Habernal2017ArgumentationMI}, the valuable auxiliary information from the other modalities is directly deserted in the construction of datasets. Furthermore, the existing corpora are relatively small in scale, and a large proportion of them do not support the subtask of relation prediction. 

\begin{figure}[ht]
\centering
\includegraphics[width=1.02\linewidth]{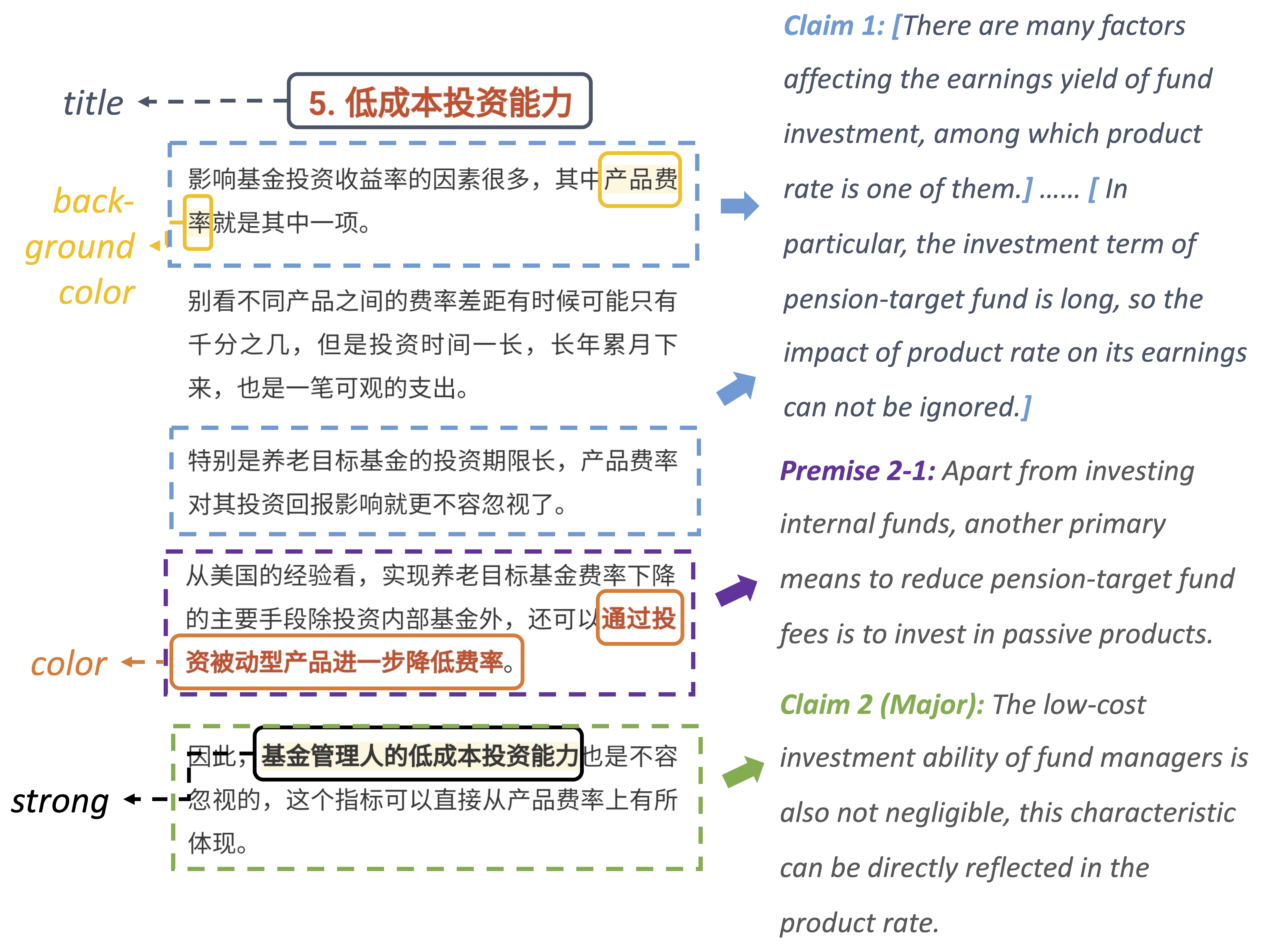}
\caption{An example of visually-rich free-form documents in our proposed \textit{AntCritic} dataset. Better viewed with color and zoom-in.}
\label{fig:intro}
\end{figure}

Considering these aspects mentioned above, we collect a novel dataset named \textit{AntCritic} to facilitate the research of argument mining on visually-rich documents and free-form expressions. Concretely, we collect a total of about 10k argumentative financial comments from an open online forum supported by \textit{Alipay}, which achieves the largest scale in this field as far as we know. In these documents, the visual patterns represented in the format of HTML tags and attributes are provided along with the text statements, and all the arguments are allowed to represent in a \textit{free} form, i.e., a semantic-consistent argument might be scattered in non-adjacent segments, which is illustrated in Figure \ref{fig:intro}. This setting broadens the original definition of this problem and paves the way for possible future research. On this basis, to provide a preliminary solution to free-form argument mining, we propose a fine-grained setting as an extension to this field and explore different predictive components to tackle the corresponding problem. As for the fusion of visual patterns, we develop a style gate module to measure the effect of style attributes and utilize a joint encoding mechanism to represent the complex positions in the layout of documents. Moreover, we also develop different model variants to conduct in-depth analysis and discussions on the granularity of modeling units. 

In conclusion, our contributions in this paper can be explained in the following three aspects.
\begin{itemize}
	\item We collect and contribute a novel dataset \textit{AntCritic} to argument mining. This corpus contains about 10k visually-rich free-form financial comments, which makes it the largest corpus in this area to our best knowledge.
	\item We thoroughly discuss the problem setting and the corresponding solution to free-form argument mining and explore a possible scheme to take advantage of auxiliary visual pattern information, which further broadens the scope of this field.
	\item We design two simple but effective model architectures and conduct extensive experiments to provide a reliable benchmark for our proposed dataset and serve as a preliminary solution to the task of argument mining on visually-rich free-form documents.  
\end{itemize}

\section{Related Work}
\subsection{Visually-rich Document Understanding}
Depending on the accessibility of data organization, visually-rich documents can be roughly categorized into two groups. The first ones are \textit{pattern-accessible} documents, such as webpages, and markdown files, in which all the visual patterns are controlled by clear scripts and can be explicitly accessed by analysts. On the contrary, in \textit{pattern-inaccessible} documents like tickets and posters, there barely exist direct clues or instructions for the construction of documents, resulting in the vagueness of visual analysis for them. In this paper, we mainly discuss the former cases because the visual attributes represented by HTML tags are actually pattern-accessible ones.

Guided by explicit visual instructions, the rendering markups can greatly help analyze documents in a structured and human-like way. 
\citet{Wang2022WebFormerTW} consider HTML tags and linguistic tokens as two individual sequences to model their relations in a cross-fusion way. \citet{Li2021MarkupLMPO, Deng2022DOMLMLG} bring the self-supervised pre-training diagram into this field to drive models comprehensively to understand the whole webpage. Apart from these, some specialized architectures are also devised for some downstream tasks: \citet{Ashby2020LeveragingHI} regard HTML tags as auxiliary information to recognize named entities. Also, \citet{Chen2021WebSRCAD} collect a novel question-answering dataset and formulate the task of structural reading comprehension on the web. Furthermore, \citet{Wang2021TCNTC} design a novel attention-based framework to generate relational table representations in finer granularity. These works inspire us to boost the performance of argument mining with the help of visual and structural information.


\subsection{Argument Mining}
Argument mining aims to automatically detect and model all possible argument structures in the given documents or text fragments. In previous studies, researchers primarily focus on classifying argument components. 
\citet{Lawrence2014MiningAF, Nguyen2016ContextawareAR} apply Latent Dirichlet Allocation (LDA) to learn in an unsupervised way. \citet{Madnani2012IdentifyingHO} employ Conditional Random Field (CRF) to make predictions based on some pre-defined lexical properties. \citet{Alhindi2021SharksAN, Zhang2022EnhancingLA} leverage the prevailing BERT and Bi-LSTM modules to tackle this problem with the help of pre-training. Operating on the constituency tree, \citet{Ruggeri2021TreeConstrainedGN} extract semantic information in a graph-based way. Moreover, \citet{Daxenberger2017WhatIT, Schulz2018MultiTaskLF} explore the transferability between different domains, and \citet{Gao2017ReinforcementLB, Kees2021ActiveLF} thoroughly discuss the feasibility of reinforcement learning and active learning in this field, respectively. Besides, \citet{Nguyen2016ContextawareAR, Peldszus2015TowardsDC, Cocarascu2018CombiningDL} also devise various relation modeling schemes to identify the relations between arguments, which further advance the identification of argumentative structure at a higher level. 

Considering it is pretty essential to provide enough samples and annotations for machine inference and reasoning, some valuable datasets are also collected from various domains, covering the types of scientific papers\cite{Mayer2020TransformerBasedAM}, persuasive essays\cite{Stab2017ParsingAS}, sustainable development policies\cite{Fergadis2021ArgumentationMI}, legal texts\cite{Zhang2022EnhancingLA} and so forth. Apart from these, \citet{Bosc2016DARTAD, Liu2017UsingAF, Habernal2018TheAR} also collect datasets containing argument structures for the recognition of emotion or opinion preferences in comments and reviews. Lately, \citet{Mestre2021MArgMA} contribute a large-scale audio-text dataset, promoting the application of multi-modal analysis in this area. However, these datasets are limited in scales and expression forms, and there is still no adequate research to combine other information into the analysis, making our research a meaningful complement to this field.

\section{Dataset}
Based on the aforementioned considerations, we aim to contribute a large-scale corpus to this field, named \textit{AntCritic}. This dataset contains about 10k Chinese financial comments collected from an open online forum supported by \textit{Alipay}, which covers the themes of fund introductions, stock market analysis, investment advice, etc. Before collecting all the original data from the online open forum, we confirmed that the users had permanently and irrevocably licensed the rights to the published content to the platform by means of the user agreement. And we have obtained the right from the platform to use the data for academic research and public release. 


\subsection{Data Annotation}  
\begin{figure}[ht]
\centering
\includegraphics[width=\linewidth]{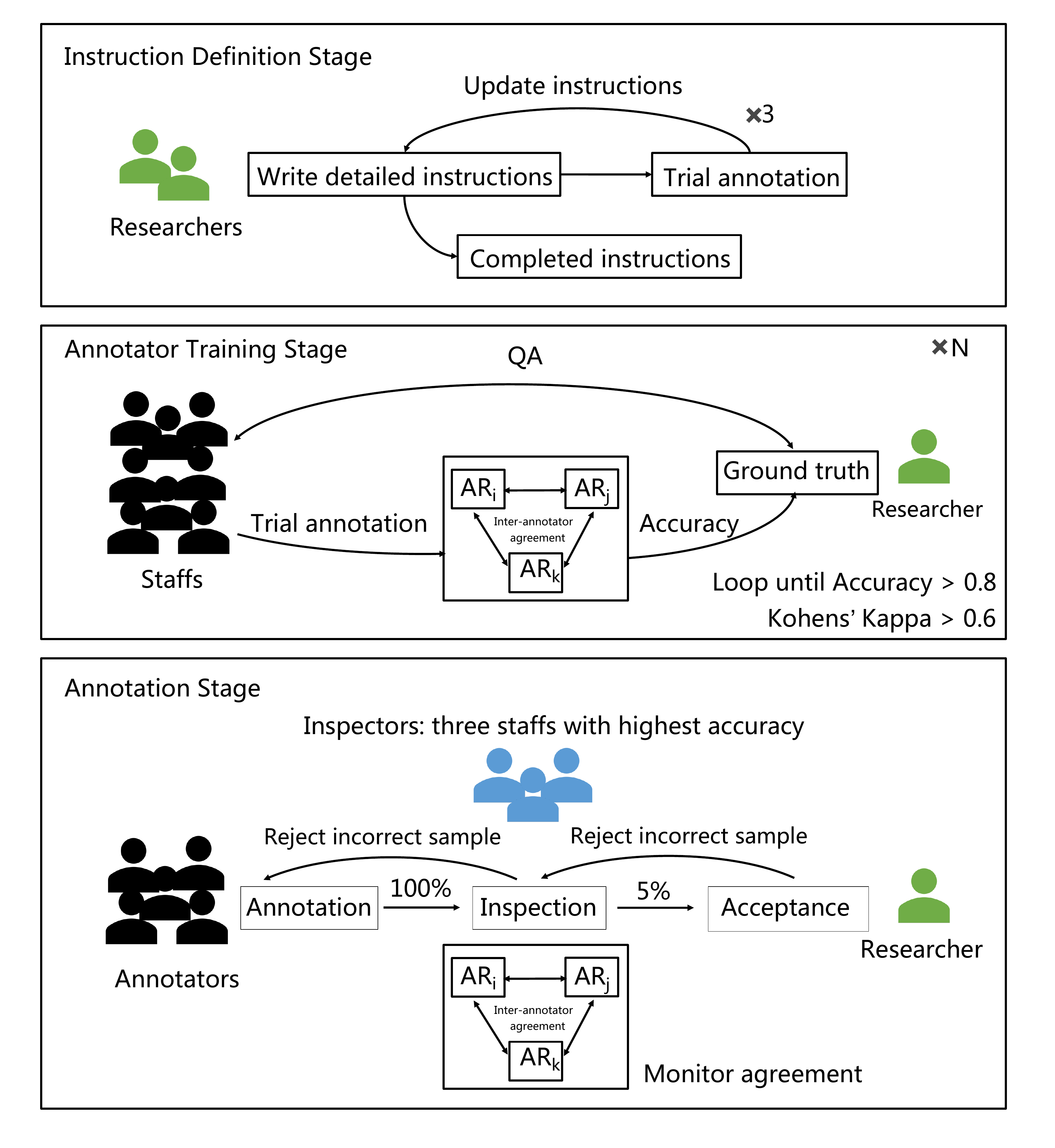}
\caption{The overall annotation pipeline.}
\label{fig:pipeline}
\end{figure}

To ensure the quality of annotated data, we develop our annotation pipeline shown in Figure \ref{fig:pipeline}.
\begin{table*}[!h]
\caption{Statistic Information Comparison of Datasets for Argument Mining}
\label{tab:datasets}
\centering
\begin{tabular}{>{\centering}m{1.3cm}|>{\centering}m{1.25cm}|c|c|c|c|c|c|>{\centering}m{1cm}|c}
\toprule \hline
 Dataset & Domain & \#Docs & \#Sents & \#Claims & \#Tokens & Unit & Rel? & Modal            & Lang                 \\ \hline
\citeauthor{Habernal2017ArgumentationMI} & Web Discourse                       & 340         & 3,899        & 211      & 84,817    & Token       & No        & Text & EN \\  \hline          
\citeauthor{Mayer2020TransformerBasedAM} & Scientific Papers  & 659 & 35,012 & 1,390 & - & Token & Yes & Text & EN \\ \hline
\citeauthor{Reed2008LanguageRF} & Various Genres                      & 507         & 2,842        & 563      & 60,383    & Clause      & No      &                    Text & EN                         \\  \hline
\citeauthor{Stab2017ParsingAS} & Persuaive Essays                    & 402         & 7,116        & 2,108     & 147,271   & Clause      & Yes                 &                    Text & EN                        \\  \hline
\citeauthor{Peldszus2015AnAC} & Micro Texts                         & 112         & 449         & 112      & 8,865     & Segment     & No                  &                 Text & EN        \\  \hline
\citeauthor{Fergadis2021ArgumentationMI} & SDG Policy & 1,000        & 12,374       & 1,202     & -        & Sentence    & No                  &                  Text & EN                       \\  \hline
 Ours & Financial Comments                  & 9,994       & 214,585      & 88,311          & 11,436,977 & Segment     & Yes                 & Text \& HTML          & CN               \\
 \hline \bottomrule 
\end{tabular}
\end{table*}
\subsubsection{Instruction Definition Stage} In this stage, we will first draw up a series of detailed instructions, including the definition and format of different related terms.  And then, we will carry out several rounds of trial annotations and refine the instructions so that the vagueness in instructions can be reduced to the minimum and the instructions are practical to all kinds of data.  And here, we list some important terms as follows.
\begin{itemize}[leftmargin=*]
	\item A \textit{claim} is defined as a group of argumentative segments that express a clear point of view, and the corresponding \textit{premises} are defined as the fragments containing factual statements which underpin this opinion. 
	\item There should be at least one claim in a document. The maximum of claims is set as 9, and the maximum of premises supporting the same claim is 4. It is also allowed to annotate some opinions as simple claims without any supporting premises.  	
	\item All the claims and premises should be semantically complete and individual. So the annotators may need to concatenate some non-adjacent segments to form a single integral argument. Besides, the annotators should also select a major claim that best matches the overall opinion. 
\end{itemize} 

\subsubsection{Annotator Training Stage} After the refinement of instructions, eight staff with specific financial knowledge are employed to annotate the collected data. Before the formal annotation, we use some examples to help them get familiar with this task's goal and the corresponding requirements. Then, we will conduct rounds of trial annotations, evaluate the performance and statistics among all the annotators, and resolves the doubts of annotators in the QA phase. This process will repeat until the average accuracy reaches 80\% and the Cohen's kappa \cite{mchugh2012interrater} agreement index is greater than 60\%. Afterward, three annotators with the highest accuracy are selected as inspectors in the next formal annotation stage.

\subsubsection{Annotation Stage}
In the final stage, the formal annotation stage will be carried out in three steps: \textit{Annotation-Inspection-Acceptance}. As mentioned above, a total of five annotators are required to label the data. And then, the other three inspectors check all the labeled results and reject the incorrect results. Finally, we sample 5\% data to monitor the overall quality of annotation, and we accept the final annotations if and only if the agreement index and accuracy can reach the requirement.

\subsection{Data Properties} 
To the best of our knowledge, our contributed \textit{AntCritic} is the largest argument mining dataset in scale, and the detailed comparison can be found in Table~\ref{tab:datasets}. Besides, there are also some unique properties compared with other datasets proposed previously, which are depicted in Figure \ref{fig:intro} and \ref{fig:tree}, and listed as follows.  \italicRoman{1} \textit{Multi-Modality}: all the visual patterns, structural information and linguistic expressions can be accessed directly and explicitly. \italicRoman{2} \textit{Discontinuity}: the text segments belonging to the same argument may be placed in a discontinuous manner.  \italicRoman{3} \textit{Non-Monotonicity}: some segments of different argumentative components may be arranged alternatively.


These characteristics described above primarily stem from the nature of the data source. Because there are no strict writing instructions for creators on the platform, the organization of documents may not be compact and precise enough, thus making the segments from different argument components interweave with each other. Although these properties significantly increase the learning difficulty, this setting reduces the requirements on the corpus to the minimum, which provides more opportunities for further research on free-form argument mining and improves the feasibility and generalization of learned models in practice.

\subsection{Dataset Statistics}
\begin{table}[ht]
  \caption{Training/Validation/Test corpora statistics.}
  \label{tab:statistics}
  \begin{tabular}{cccc}
    \toprule
     &Train & Validation & Test\\
    \midrule
    \# Comments & 7,986 & 1,007 & 1,001 \\
    \# Segments & 433,342 & 40,296 & 33,913\\ 
    \# Claims & 35,036 & 4,741 & 4,065\\ 
    \# Premises & 37,022 & 4,102 & 3,788\\ \hline
    \# \makecell{Claim \\  Segment} & 69,823 & 10,225 & 8,263\\ \hline
    \# \makecell{Premise \\ Segment} & 127,379 & 13,620 & 11,321\\ \hline
    \makecell{\# Character \\ / Segment} & 22.55 & 22.10 & 22.84\\ \hline
    \makecell{\# Segment \\ / Claim} & 1.99 & 2.16 & 2.03\\ \hline
    \makecell{\# Segment \\ / Premise} & 3.44 & 3.32 & 2.99\\ 
    \toprule
    \% Font & 16.85\% & 10.73\% & 18.72\%\\
    \% Strong & 9.98\% & 7.30\% & 10.10\%\\
    \% Color & 7.07\% & 6.03\% & 9.37\%\\
    \% Blockquote & 0.97\% & 0.62\% & 0.58\%\\
    \% Supertalk & 0.64\% & 0.69\% & 0.96\%\\
    \% Header & 0.24\% & 0.59\% & 0.78\%\\
  \bottomrule
\end{tabular}
\end{table}

Table \ref{tab:statistics} demonstrates the statistic information of our proposed \textit{AntCritic}. The upper part demonstrates the quantitative characteristics of texts, and the lower part shows the percentage of segments containing specified HTML tags. 

\section{Problem Definition}
From an overall perspective, our goal of argument mining on visually-rich free-form documents can be decomposed into two targets, namely \textit{Argument Component Detection} and \textit{Argument Relation Prediction}, formulated as follows.

\begin{figure}[h]
\centering
\includegraphics[width=\linewidth]{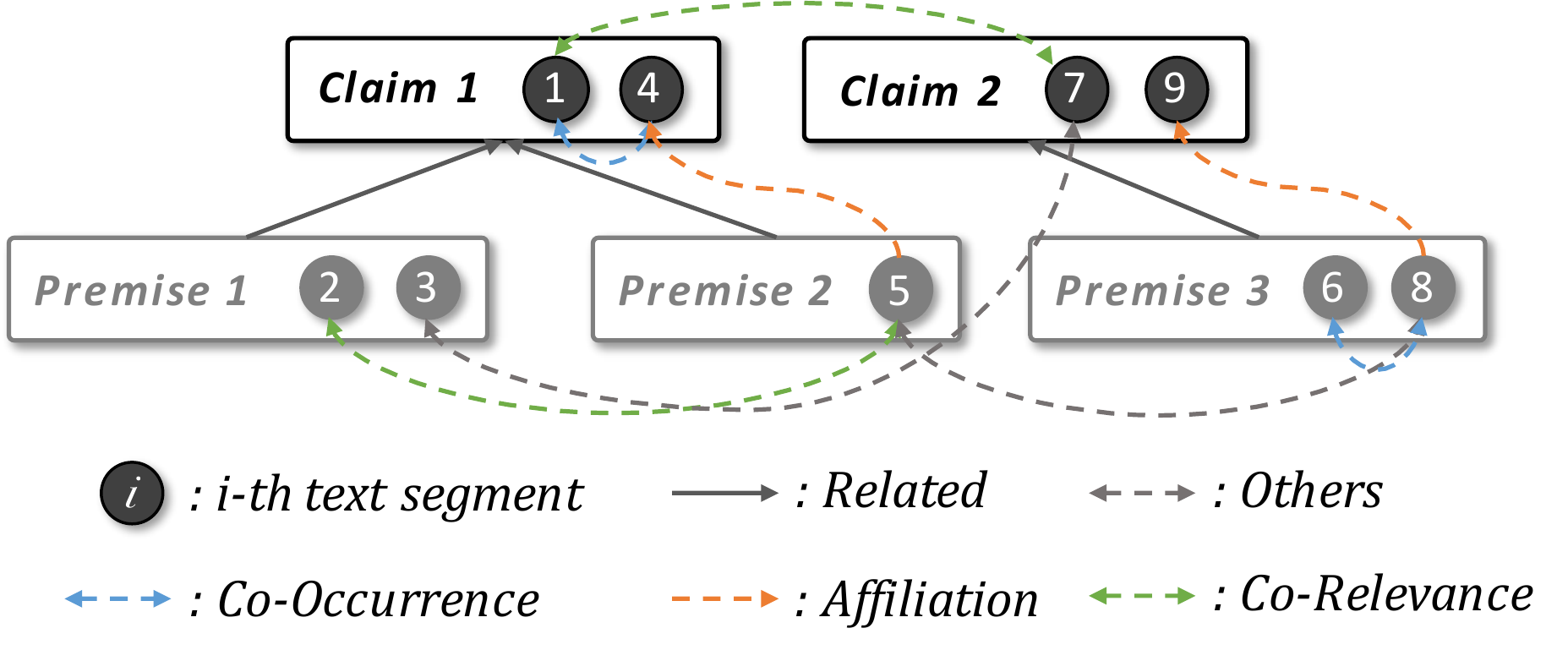}
\caption{A simple diagram example for argument structures. Better viewed in color. }
\label{fig:tree}
\end{figure}

\subsection{Argument Component Detection} As the crucial step in the task of argument mining, it aims to detect and classify all the potential argument components. Based on the annotation of this dataset, we only consider three possible component types, including \textit{Claim}, \textit{Premise} and \textit{Non-Argument}. Apart from this, the confidence of arguments to be major claims should also get estimated to serve as a reference for human decision-making. 

\subsection{Argument Relation Prediction} To recover the argument structure in the documents, the relations between arguments should be understood correctly. However, the \textit{discontinuity} and \textit{non-monotonicity} of the dataset heavily impede us from slicing the entire document into multiple consecutive parts and predicting their relations. In light of this, we turn to recognize the relation types between individual segments and construct a more dense relation network within documents. Therefore we sort out all possible situations and accordingly define four relation types to cover them.
\begin{itemize}[leftmargin=*]
	\item \textbf{Affiliation: } The $i$-th segment is defined to \textit{affiliate} with the $j$-th one if the latter belongs to a claim and the former is part of the premise supporting this claim. It maintains the vertical hierarchy of the entire document.
	\item \textbf{Co-Occurrence: } The $i$-th and $j$-th segments are defined to form a \textit{Co-Occurrence} relation if they belong to the same argument. Without loss of generalization, every argumentative segment builds a \textit{co-occurrence} relation with itself. It reveals the internal composition of a single argument component.
	\item \textbf{Co-Relevance: } The $i$-th and $j$-th segments are defined to form a \textit{Co-Relevance} relation if they belong to the sibling components which support the same argument. And we assume that all the claims are trivially defined to support a virtual main topic. This kind of relationship can help recover the horizontal argumentative structure of the document.
	\item \textbf{Other: } The \textit{Other} relation includes all the cases not covered by the definitions listed above.
\end{itemize} 
Their diagram can also be found in Figure \ref{fig:tree}. It is straightforward to notice that there tend to be some redundancies in the relation annotations. 
These redundancies seem unnecessary, but they can enhance the robustness of relation predictions and act as an extra constraint to guide the model to consider the global argumentative structure and understand the relation types in depth. 

\section{Method}
\subsection{Model Architecture}
This section will describe our proposed preliminary solution to the aforementioned problem setting on the \textit{AntCritic} dataset. Given a structured webpage with various markups and attributes $\mathbf{D} = \{(\mathbf{d}_i, \mathbf{m}_i, p_i, s_i)\}_{i=1}^n$, our goal is to predict the labels of text segments $\mathbf{L} = \{l_i\}_{i=1}^n$, the confidence to be major claims $\{c_i\}_{i=1}^n$, and the relations between them $\mathbf{R} = \{\{r_{ij}\}_{i=1}^n\}_{j=1}^n$, where $n$ is the total number of text segments and the quadruple $(\mathbf{d}_i, \mathbf{m}_i, p_i, s_i)$ represents the token sequence, style marks, paragraph position and segment position of the $i$-th text segment respectively. To quantify the impact of modeling granularities on the prediction, we construct the token-level and segment-level architectures to extract information and make predictions individually.
 
\subsubsection{Token-Level Model}
Because of the exceeding number of input tokens, we process each segment separately and only conduct the argument component detection task in this part. The overall pipeline of this model is demonstrated in the appendix. We consider both the character and word sequences of the input segments to investigate all the possible modeling schemas and further explore the dataset's properties.
Concretely, we extract the aggregated representations via character-based and word-based backbones in a \textit{First-Last-Avg} manner and finetune these models. Afterward, we apply a simple multi-layer perceptron to generate class probabilities. And following the approach proposed by \citet{Wei2019EDAED}, three kinds of strategies for text augmentation are also employed to enhance the robustness of predictions, namely \textit{Random Mask}, \textit{Random Swap} and \textit{Random Repeat}. In this computing process, we leave out the HTML tags from the calculation because the marks for a single segment can only provide quite limited structural information and visual difference for the label prediction. 


\subsubsection{Segment-Level Model} 
In the initialization, we also adopt the \textit{First-Last-Avg} mechanism to generate the segment-level features, but all the parameters in backbones are frozen for less computation resource requirements.

Afterward, we project these features into the same latent space, denoted as $\mathbf{f}^\mathrm{c}_i$ and $\mathbf{f}^\mathrm{w}_i$ for the character-base and word-based aggregations of the $i$-th segment, respectively. Next, considering that these two groups of embeddings are complementary semantic views of the input segments that emphasize different semantic aspects, we fuse these two representations into a single feature sequence. To be more specific, we apply a cross-gate module to calculate the fused representations adaptively. The formulae are given as
\begin{equation}
\label{eq:cross}
\begin{array}{lr}
\mathbf{g}^\mathrm{c}_i = \sigma(\mathbf{W}^\mathrm{c}\mathbf{f}^\mathrm{c}_i + \mathbf{b}^\mathrm{c}), & \mathbf{g}^\mathrm{w}_i = \sigma(\mathbf{W}^\mathrm{w}\mathbf{f}^\mathrm{w}_i + \mathbf{b}^\mathrm{w}),
\end{array}
\end{equation}
\begin{equation}
\mathbf{f}_i = \mathbf{W}^\mathrm{o} \cdot [\mathbf{g}^\mathrm{c}_i \cdot \mathbf{f}^\mathrm{w}_i;\mathbf{g}^\mathrm{w}_i \cdot \mathbf{f}^\mathrm{c}_i] + \mathbf{b}^\mathrm{o},
\end{equation}

where $\mathbf{g}^*_i \in (0,1)^d, \mathbf{f}^*_i \in \mathbb{R}^{d}$ and the variables $\mathbf{b}^* \in \mathbb{R}^{d}$, $\mathbf{W}^\mathrm{w}, \mathbf{W}^\mathrm{c} \in \mathbb{R}^{d \times d}$, $\mathbf{W}^\mathrm{o} \in \mathbb{R}^{d \times 2d}$ are all learnable parameters. $\sigma(\cdot)$ is the sigmoid function and $[;]$ is the concatenation operator. 


Meanwhile, the visual pattern $\mathbf{m}_i$ and structural position information $(p_i, s_i)$ are also encoded into a series of embeddings denoted as $\mathbf{v}_i$ and $\mathbf{e}_i$, which can be given by
\begin{equation}
\begin{array}{lr}
\mathbf{e}_i = [\mathbf{E}^\mathrm{p}_{p_i}; \mathbf{E}^\mathrm{s}_{s_i}], &
\mathbf{v}_i = \mathbf{E}^\mathrm{m}\mathbf{m}_i,
\end{array}
\end{equation}
where $\mathbf{m}_i \in \{0,1\}^6$ is the stack of indicators for font, strong, color, blockquote, supertalk and header tags / attributes in the $i$-th segment, respectively. And $\mathbf{E}^\mathrm{p}, \mathbf{E}^\mathrm{s} \in \mathbb{R}^{N \times (d/2)}, \mathbf{E}^\mathrm{m} \in \mathbb{R}^{d \times 6}$ are the trainable weights of look-up tables.

Then, we utilize a style gate to dynamically control the density of semantic information according to the style appearances and reveal what the authors want to emphasize. The pattern-aware embedding $\{\mathbf{\hat{f}}_i\}_{i=1}^n$ will be consequently generated as
\begin{equation}
\begin{array}{lr}
	\mathbf{g}^\mathrm{g}_i = \sigma(\mathbf{W}^\mathrm{g}[\mathbf{v}_i;\mathbf{f}_i] + \mathbf{b}^\mathrm{g}), & \mathbf{\hat{f}}_i = (1 + \mathbf{g}^\mathrm{g}_i)\mathbf{f}_i,
\end{array}
\end{equation}
where $\mathbf{g}^\mathrm{g}_i \in (0,1)^d, \mathbf{\hat{f}}_i \in \mathbb{R}^d$, and $\mathbf{W}^\mathrm{g} \in \mathbb{R}^{d \times 2d}, \mathbf{b}^\mathrm{g} \in \mathbb{R}^d$ are learnable parameters. 

And next, we need to utilize a sequence model to integrate the global information and form a series of context-aware representations given by
\begin{equation}
(\mathbf{\tilde{f}}_1, \dots, \mathbf{\tilde{f}}_n) = \Omega((\mathbf{\hat{f}}_1 + \mathbf{e}_1),  \dots, (\mathbf{\hat{f}}_n+\mathbf{e}_n); \Theta),
\end{equation}
where $\Omega(\cdot; \Theta)$ denotes the sequence modeling component. In practice, we select Transformer \cite{Vaswani2017AttentionIA} and GRU \cite{Chung2014EmpiricalEO} as typical choices in the experiment. 

Consequently, context-aware representations will be used to detect all the arguments and recognize the relations between segment pairs. For argument component detection, we employ two linear layers to obtain the class probabilities $\mathbf{a}_i \in (0,1)^3$ and major confidence $c_i \in (0,1)$, given by
\begin{equation}
\begin{array}{lr}
	\mathbf{a}_i = \phi(\mathbf{W}^\mathrm{a} \mathbf{\tilde{f}}_i + \mathbf{b}^\mathrm{a}), &
	c_i = \sigma(\mathbf{W}^\mathrm{m}\mathbf{\tilde{f}}_i + b^\mathrm{m}),
\end{array}
\end{equation}
where $\mathbf{W}^\mathrm{a} \in \mathbb{R}^{3 \times d}, \mathbf{W}^\mathrm{m} \in \mathbb{R}^d, \mathbf{b}^\mathrm{a} \in \mathbb{R}^{3}, b^\mathrm{m} \in \mathbb{R}$ and $\phi(\cdot)$ is the softmax operator. 

And for the task of argument relation prediction, we first project the features into the subspaces corresponding to the source and destination ends of relations, given as
\begin{equation}
		\mathbf{f}^*_i = \mathbf{W}^*_2(\operatorname{ReLU}(\mathbf{W}^*_1\mathbf{f}_i+\mathbf{b}^*_1)) + \mathbf{b}^*_2, \, * \in \{s, d\}.
\end{equation}
where the dimension of parameters keeps consistent with Equation \ref{eq:cross}. Next, two different pair-wise classification components are utilized to calculate the relation probabilities $\mathbf{r}_{ij} \in (0,1)^4$, which can be formulated as
\begin{itemize}[leftmargin=*]
	\item \textbf{Mul-Add: }
	\begin{equation}
	\label{eq:mul-add}
		\mathbf{r}_{ij} = \phi(\mathbf{W}^\mathrm{r}[\mathbf{f}^\mathbf{s}_i + \mathbf{f}^\mathbf{d}_j; \mathbf{f}^\mathbf{s}_i \cdot \mathbf{f}^\mathbf{d}_j]+\mathbf{b}^\mathrm{r}),
	\end{equation}
	\item \textbf{Biaffine: }
	\begin{equation}
	\label{eq:biaffine}
		\mathbf{r}_{ij} = \phi({[\mathbf{f}^\mathbf{s}_i};1]^\mathsf{T} \mathbf{U} [\mathbf{f}^\mathbf{d}_j;1]+\mathbf{b}^\mathbf{r}),
	\end{equation}
\end{itemize}
where $\mathbf{U} \in \mathbb{R}^{(d + 1) \times 4 \times (d + 1)}, \mathbf{W}^\mathrm{r} \in \mathbb{R}^{4 \times 2d}, \mathbf{b}^\mathrm{r} \in \mathbb{R}^d$ are trainable parameters. And the detailed calculation mechanism of the segment-level model is shown in the appendix.

\subsection{Training and Inference}
As illustrated in the descriptions in the sections above, the overall calculation can be conducted end-to-end. Considering the entire task is composed of three subtasks (including major confidence estimation), we optimize the model by combining these three objectives, formulated as follows.

\paragraph{Argument Component Detection }The subtask of argument component detection is essentially a multi-classification problem, so we apply the cross-entropy loss function to tackle this.
	\begin{equation}
			\mathcal{L}_c = -\sum_{i=1}^{n} \sum_{j=0}^2 \mathbb{I}(\hat{l}_i = j) \operatorname{log} ((\mathbf{a}_i)_j),
	\end{equation}
	where the ground-truth component label $\hat{l}_i$ equals to 0, 1, 2 if the $i$-th segment is annotated as \textit{non-argument}, \textit{claim} and \textit{premise}, respectively.

\paragraph{Argument Relation Prediction } Similarly, the optimization constraint of argument relation prediction can be given by
	\begin{equation}
			\mathcal{L}_r = -\sum_{i=1}^{n} \sum_{j=1}^{n} \sum_{k=0}^3 \mathbb{I}(\hat{r}_{ij} = k) \operatorname{log} ((\mathbf{r}_{ij})_k)
	\end{equation}
	where the ground-truth relation label $\hat{r}_{ij}$ is assigned as 0, 1, 2, 3 where the relation between $i$-th and $j$-th segments is considered as \textit{other}, \textit{affiliation}, \textit{co-occurrence} and \textit{co-relevance}, respectively.

\paragraph{Major Confidence Estimation } The estimation of major confidence will be approximated as a binary classification task and get solved by
\begin{equation}
\mathcal{L}_m = -\sum_{i=1}^{n} (\hat{m}_i \operatorname{log}(c_i) + (1 - \hat{m}_i) \operatorname{log}(1 - c_i)),
\end{equation}
where $\hat{m}_i = 1$ if the $i$-th segment is annotated as a part of the major claim, otherwise $\hat{m}_i = 0$.
 
Finally, the segment-level model will be optimized in a multi-task manner, and the overall loss function is given by
\begin{equation}
	\mathcal{L} = \lambda_c\mathcal{L}_c + \lambda_r\mathcal{L}_r + \lambda_m\mathcal{L}_m,
\end{equation}
where $\lambda_*$ are the balancing hyper-parameters. As for the token-level model, we only calculate the first term because the other two subtasks are not conducted on this. 

While in the inference, we directly output the major confidence $c_i$ and choose the labels with the highest probabilities as the prediction results.


\section{Experiment}
\begin{table*}[ht]
\caption{Results of different settings for segment-level models. The best results are given in \textbf{bold}.}
\centering
\label{tab:second}
\begin{tabular}{cccc|cccc|ccc}
\toprule \hline
\multirow{2}{*}{Row} & \multirow{2}{*}{\makecell[c]{Module}} & \multirow{2}{*}{\makecell[c]{Relation \\ Modeling}} & 
\multirow{2}{*}{\makecell[c]{Using \\ HTML?}} & \multicolumn{4}{c|}{Component Detection}                  & \multicolumn{3}{c}{Relation Prediction} \\ 
           &			&                &                           & Mac.       & Mic.       & Weig.    & Major     & Mac.       & Mic.       & Weig.    \\ \hline
1	&	MLP      & Biaffine               & Yes              & 63.24          & 66.68          & 66.35          & 43.27          & 20.26          & 26.05          & 20.64          \\
2	&	MLP      & Mul-Add               & Yes              & 62.59          & 66.07          & 65.75          & 42.29          & 22.61          & 27.18          & 22.93         \\
3	&	TransFM  & Biaffine              & Yes              & 65.78          & 69.01          & 68.79          & 47.49          & 21.64          & 26.34          & 22.27          \\
4	&	TransFM   & Mul-Add             & No              & 65.16          & 68.49          & 68.10          & 46.01          & 21.60          & 26.55          & 22.94          \\
5	&	TransFM  & Mul-Add              & Yes              & 65.24          & 68.67          & 68.28          & 46.30          & \textbf{23.84}          & 28.58          & 24.11          \\
6	&	Bi-GRU          & Biaffine              & Yes              & \textbf{67.23} & \textbf{70.22} & \textbf{70.02} & 48.75 & 21.66 & 26.27          & 21.68 \\
7	&	Bi-GRU          & Mul-Add              & No                      & 66.24         & 69.72          & 69.16          & 49.28          & 22.00          & 27.55 & 23.03          \\
8	&	Bi-GRU          & Mul-Add              & Yes              & 67.05 & 70.00 & 69.84 & \textbf{49.88} & 23.49 & \textbf{29.84}          & \textbf{24.77} \\
\hline \bottomrule 
\end{tabular}
\end{table*} 
\subsection{Metrics}
We adopt three trustworthy criteria to evaluate our proposed solutions, namely \textit{Component F1 Score}, \textit{Relation F1 Score} and \textit{Major Density}. For the Component \& Relation F1 Scores, we calculate the \textit{micro-}, \textit{macro-} and \textit{weighted-}F1 scores on the results of argument component detection and relation prediction. It's worth noting that the \textit{Other} type of relation will not be considered in the metric calculation because this type does not mean that there is \textit{no} relationship between the corresponding segments.
As for the \textit{Major Density}, we choose to normalize all the confidences of claim segments and sum up the scores corresponding to the ground-truth major claims, which can be given by
\begin{equation}
M = \frac{\sum_{i=1}^n(\tilde{c}_i \cdot \mathbb{I}(\hat{m}_i = 1))}{\sum_{j=1}^n\tilde{c}_j}
\end{equation}
in which $\tilde{c}_i=c_i$ when $\hat{l}_i = 1$ and otherwise $\tilde{c}_i=0$. In this mechanism, we can quantify the relative confidence density of major claims without being affected by the ratio of major segments.

%


\subsection{Comparison and Analysis}
In this section, we conduct a series of experiments to estimate the performance of the token-level and segment-level models and provide benchmark results on our proposed \textit{AntCritic} dataset. 

\subsubsection{Analysis of Token-Level Model}
The performances of different settings for the token-level model are listed in Table \ref{tab:first}. The first two rows display the performances of word-based and character-based pipelines, respectively. And in the \textit{Ensemble} setting, the final outputs are given as the average results of these two streams. From the table, we can find that the \textit{Ensemble} setting outweighs the others in general, and the word-based stream is inferior to the character-based one. This can be explained in the following three folds. First, the character-based backbone is trained on the financial corpus, and the word-based one is trained on the data of Wikipedia and news commentaries. There will be a larger domain and semantic gap for the word-based backbone to fit our proposed dataset. Second, the word vocabulary is much bigger than the character one, making the token embeddings of some low-frequency words not well-tuned. Besides, the ensemble strategy can reduce the vagueness and uncertainty of probabilities and enhance the stability of predictions.

\begin{table}[ht]
\caption{Results of different settings for token-level models. The best results are given in \textbf{bold}.}
\label{tab:first}
\centering
\begin{tabular}{c|ccc}
\toprule \hline
Setting      & Macro & Micro & Weighted \\ \hline
Word      & 54.30 & 61.64 & 58.73    \\
Character & 57.62 & 63.39 & 61.56    \\
Ensemble  & \textbf{58.32} & \textbf{63.87} & \textbf{62.15}    \\
\hline \bottomrule
\end{tabular}
\end{table}

\subsubsection{Analysis of Segment-Level Model} Table \ref{tab:second} demonstrates the evaluation results corresponding to different components and input modalities choices. And we try to analyze the performances in the following three aspects.

	\paragraph{Impact of Modal Inputs} As shown in Row 4,5,7,8 of Table \ref{tab:second}, it is evident that the visual patterns and structural information indeed improve the performances of all these tasks to different extents, which implies that the authors and creators tend to utilize some special attributes and layouts to emphasize their opinions and the corresponding expressions, and our proposed gating mechanism can capture this auxiliary information and help the model to focus on both semantic information and visual appearances of segments. 

	\paragraph{Choice of Sequence Modeling Components} Both Row 1,3,6, and Row 2,5,8 illustrate the effects of different sequential modules. It is explicit that the Bi-GRU module is superior to the Transformer, and both of them achieve better performances than the simple MLP. We intuitively speculate that the range of dependency modeling may lead to the performance gap between different modules. Considering the related claims and premises are usually close in position, we infer that the Bi-GRU module keeps semantic dependencies within a relatively short term, which provides enough information for message passing and aggregation and prevents long-term noise in the calculation.



	\paragraph{Choice of Relation Predictors} Comparing Row 1,3,6 with Row 2,5,8, we can find that the choices of relation modules bring a noticeable impact on relation prediction, and slightly affect the performance on argument component detection. The \textit{Biaffine} method generally behaves worse than the \textit{Mul-Add} strategy. The reason can be inferred that the inner-product of the biaffine module mixes the information along the embedding dimension, which may smooth the differences within it. However, the operation of \textit{Mul-Add} individually models the weights of different positions and maintains the discrepancy information on this dimension. 


\subsubsection{Comparison between Different Models} Comprehensively comparing the evaluation results shown in Table \ref{tab:second} and \ref{tab:first}, we can notice that the interaction between segments brings a more significant effect than the dependency modeling within a single segment. This result aligns with the intuition that the context semantic information is more important than the inner structure of expression for the reasoning and detecting argument components.

\section{Conclusion}
In this paper, we discuss a meaningful problem extension in the field of argument mining, i.e., the processing of visually-rich free-form documents, and explore the corresponding preliminary solutions to utilize auxiliary visual information and generate free-form fine-grained predictions. Besides, we further collect and contribute a large-scale corpus \textit{AntCritic} to facilitate this setting. The comprehensive experiments verify the reliability of this dataset and the feasibility of our proposed architectures.

\section*{Acknowledgements}

This work was supported by the Ant Group Research Fund.


\section{Bibliographical References}\label{sec:reference}
\bibliographystyle{lrec-coling2024-natbib}
\bibliography{anthology,ref}

\appendix

\begin{figure*}[ht]
\centering
\includegraphics[width=\linewidth]{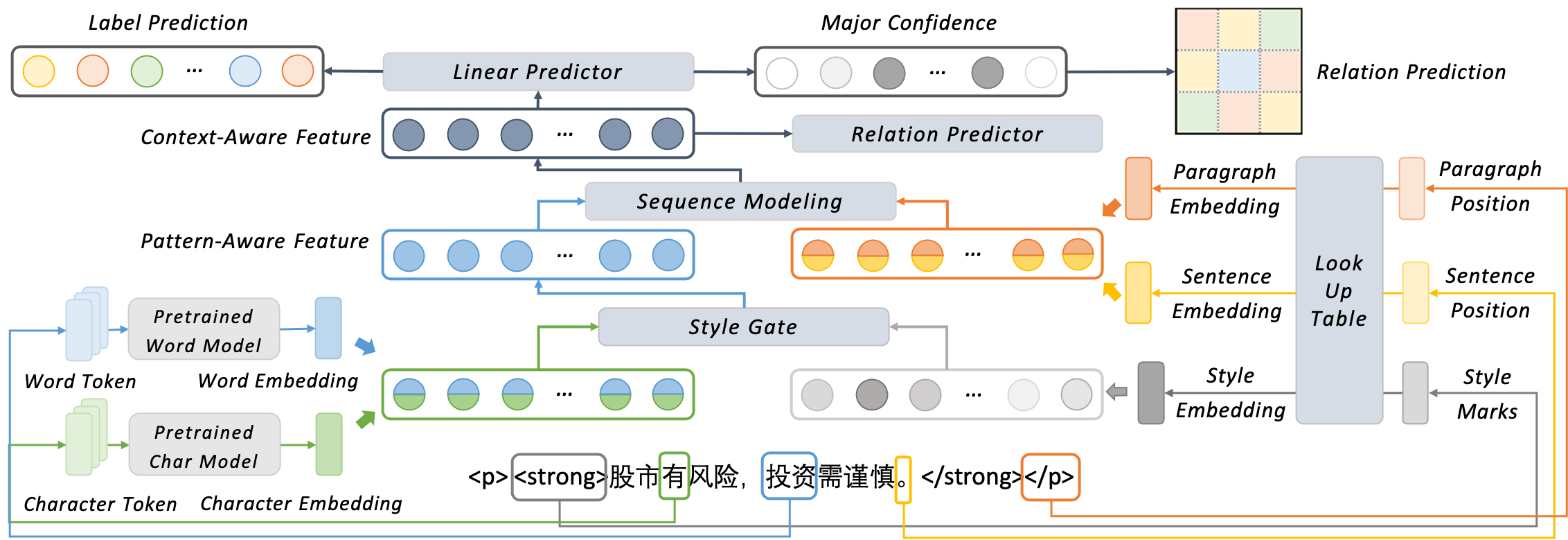}
\caption{Overall diagram of segment-level model. Better viewed in color.}
\label{fig:second}
\end{figure*}

\section{Implementation Details}
We employ the \textit{Sentence-BERT} architecture proposed by \citet{Reimers2020MakingMS} as the word-based backbone. It is constructed on the basis of BERT-like architectures and gets pre-trained on the dataset of BUCC mining task \cite{Zweigenbaum2017OverviewOT}, which is composed of Wikipedia articles and news commentaries. And the character-based backbone is adapted from \textit{FinBERT} \cite{FinBERT}, an open-source Chinese BERT model released by the AI lab of Value Simplex. The corpus used to pre-train this model is collected from the finance domain, including financial news, research reports \& announcements, and financial Wikipedia entries. The maximal number of segments $N$ in a single document is set to be 400. Except for the parameters in the pre-defined structures, the dimension $d$ of intermediate representations and learnable parameters are set to be 384. The number of layers in the Transformer and GRU components is 3, the number of heads in the attention mechanism is 4, and the layer number of the final MLP-based label predictor is set as 3. In the training process, the maximum of learning rate is set to be 1e-4, and the balancing hyper-parameters $\lambda_c, \lambda_r, \lambda_m$ are set as 1, 1 and 0.5, respectively. For both token-level and segment-level models, the learning rate is tuned with warm-up mechanism \cite{Vaswani2017AttentionIA} and cosine annealing strategy \cite{Loshchilov2017SGDRSG} where the warm-up epoch is set as 1. To prevent overfitting, a dropout strategy with $p = 0.4$ is applied in the model. The training will last for 15 epochs and we select the checkpoint with the best performance on the validation dataset. As for the optimizer configuration, the AdamW \cite{Loshchilov2019DecoupledWD, Reddi2018OnTC} optimizer with weight decay of 5e-5 is employed. The batch size is fixed to be 16. Apart from these, the probability of data augmentation for the token-level model is set as 0.3 and the ratio of modified tokens is 0.15. And in the overall experiment, the random seed is fixed and the details can be referred in the code files. 

\section{Limitation and Potential Risks}
This section will discuss the limitations and potential risks related to our proposed dataset and architecture.   In the overall view, we extend the original problem of argument mining and contribute a novel dataset to facilitate this setting.   But because there is a lack of such discussion in the prior works, we fail to conduct some transfer/transplant experiments in this problem setting.   Besides, this dataset is constructed in Mandarin, which may impede the researchers or developers in other countries from understanding and utilizing this corpus.   As for the potential risks, the inherent bias in data collection may affect the further development of this dataset.

\section{Architecture}
\begin{figure}[h]
\centering
\includegraphics[width=\linewidth]{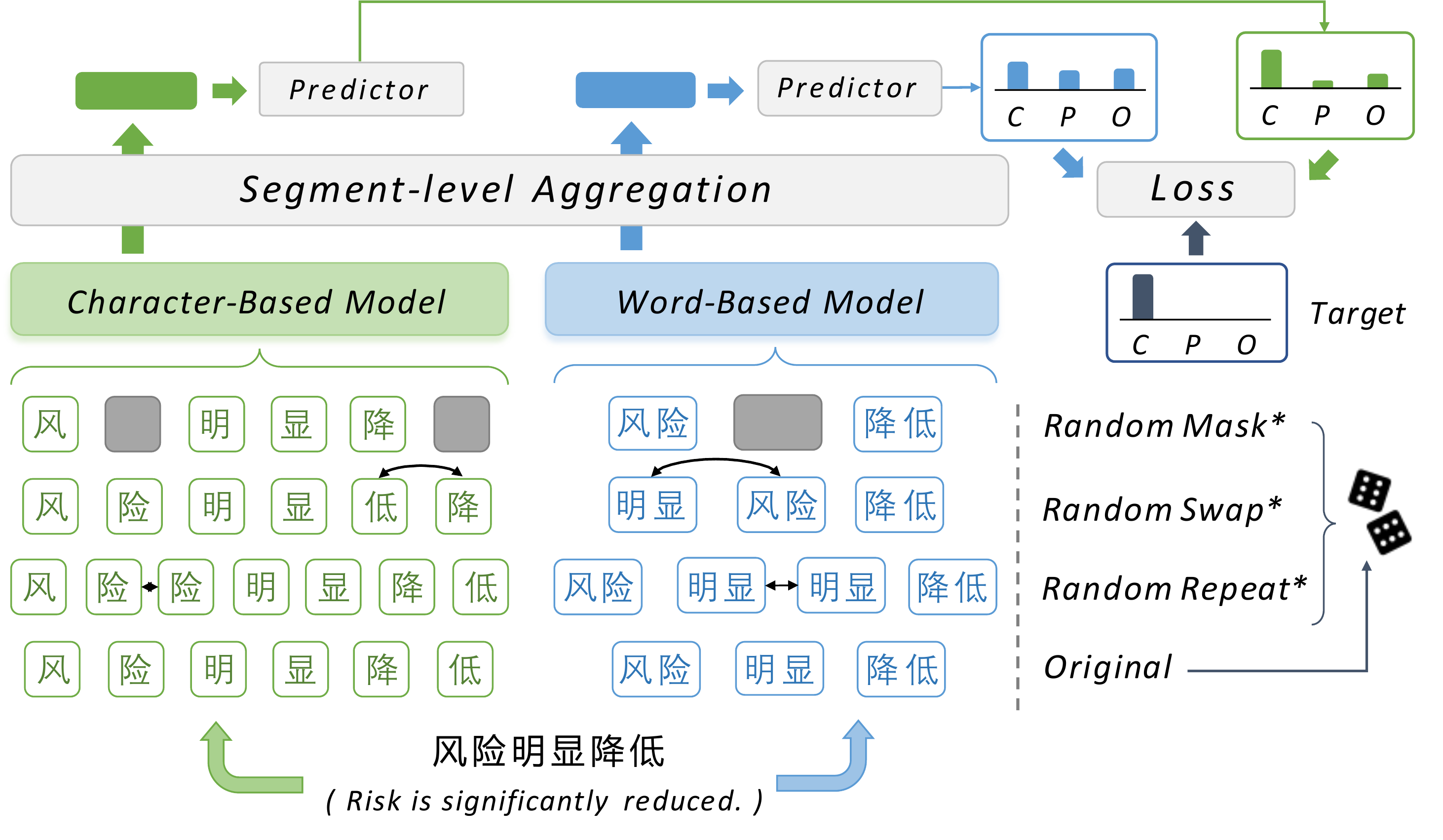}
\caption{Overall diagram of the token-level model.}
\label{fig:first}
\end{figure}

We attach our segment-level model and token-level model architecture in Figure ~\ref{fig:second} and ~\ref{fig:first}.

\section{More Dataset Samples}
We attach more visual-rich free-form documents examples here.
\begin{figure*}[h]
\centering
\includegraphics[width=\linewidth]{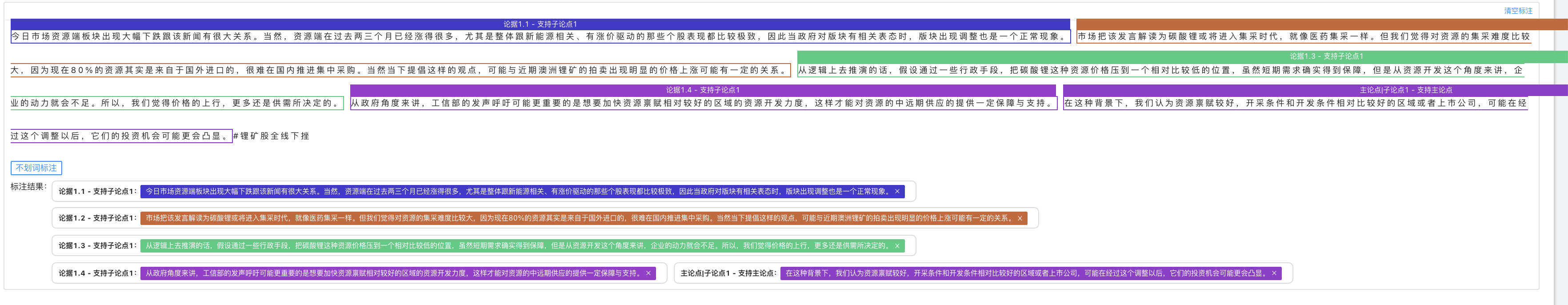}
\caption{Example 1}
\end{figure*}

\begin{figure*}[h]
\centering
\includegraphics[width=\linewidth]{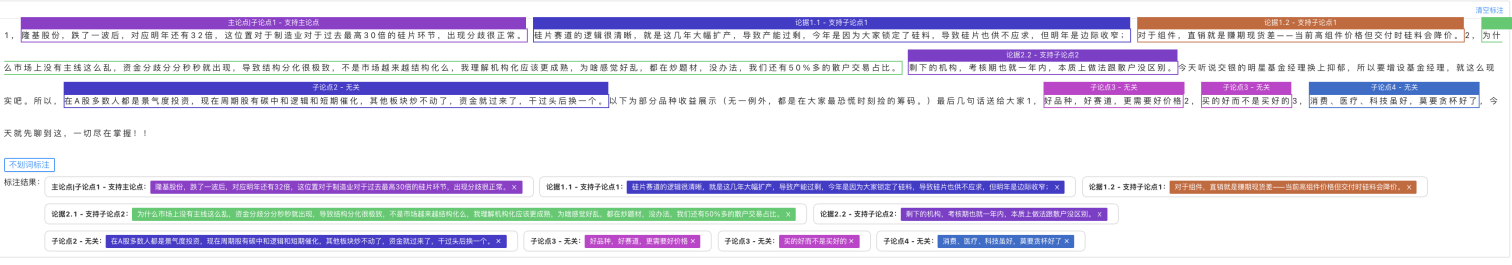}
\caption{Example 2}
\end{figure*}

\begin{figure*}[h]
\centering
\includegraphics[width=\linewidth]{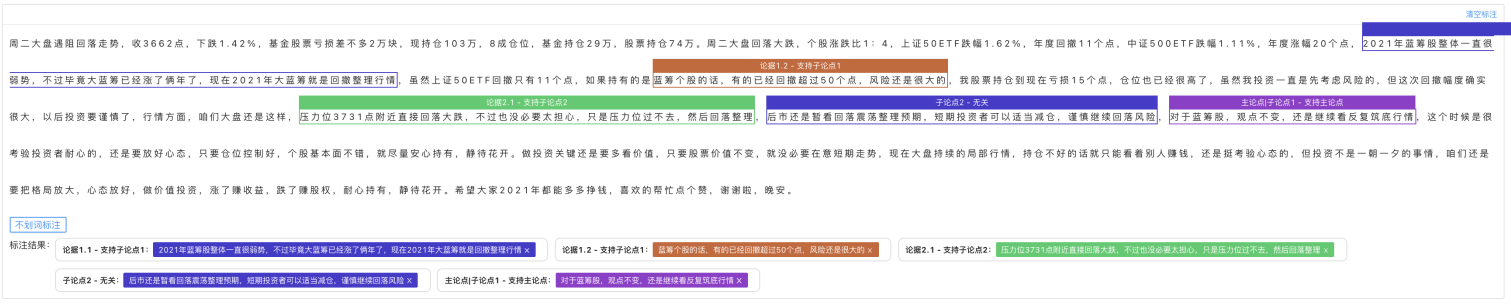}
\caption{Example 3}
\end{figure*}

\begin{figure*}[h]
\centering
\includegraphics[width=\linewidth]{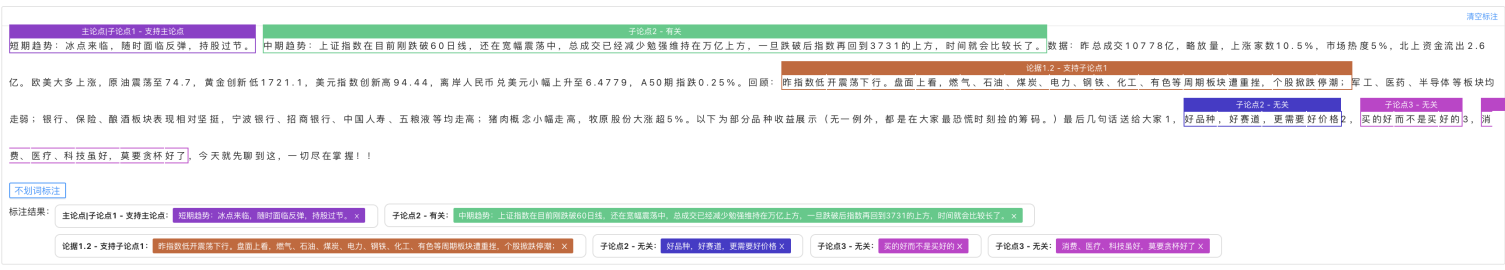}
\caption{Example 4}
\end{figure*}

\end{document}